\documentstyle[amstex,12pt]{article}                 
\newlength{\dinwidth}
\newlength{\dinmargin}
\newcommand{\CC}{{\mbox{{\small $ \Bbb C$}}}}
\newcommand{\RR}{{\mbox{{\small $ \Bbb R \,$}}}}

\newcommand{\ba}{{\mbox{{\small \boldmath $a$}}}}
\newcommand{\bp}{{\mbox{{\small \boldmath $p$}}}}

\newcommand{\bx}{{\mbox{{\small \boldmath $x$}}}}
\newcommand{\by}{{\mbox{{\small \boldmath $y$}}}}
\newcommand{\CA}{{\cal A}}
\newcommand{\CE}{{\cal E}}
\newcommand{\CF}{{\cal F}}
\def\qbox#1{\quad\hbox{#1}\quad}
\begin{document}
\title{Quarks, Gluons, Colour: Facts or Fiction?}
\author{Detlev Buchholz\\
II. Institut f\"ur Theoretische Physik,
Universit\"at Hamburg\\
Luruper Chaussee 149, D-22761 Hamburg, Germany}
\date{}
\maketitle
%
%
%
\begin{abstract}
A general method is presented which allows
one to determine from the local
gauge invariant observables of a quantum field theory the underlying
particle and symmetry structures appearing
at the lower (ultraviolet) end of the spatio--temporal scale.
Particles which are confined to small scales, i.e., do not appear
in the physical spectrum, can be uncovered
in this way without taking recourse to gauge fields
or indefinite metric spaces. In this way notions such as quark,
gluon, colour symmetry and confinement acquire
a new and intrinsic meaning which
is stable under gauge or duality transformations.
The method is illustrated by the example of the Schwinger model.
\end{abstract}
\section{Introduction}
\setcounter{equation}{0}
The structure of hadronic matter at small spatio--temporal
scales is commonly interpreted
in terms of charged particle--like structures
(quarks, gluons) which do not appear
at large scales because of confinement.
The origin of this interpretation is (a) the quark model,
which explains the form of the hadronic spectrum,
(b) the phenomenological parton picture of deep inelastic collision
processes and (c) the perturbative treatment of quantum
chromodynamics, where the particle interpretation
enters through the use of Feynman rules for the gauge
and matter fields.

In view of recent interesting developments
in gauge theory (cf.\ \cite{Se} for a review) one may,
however, be tempted to question the basis for such
an interpretation.
For in a large class of four dimensional (supersymmetric)
gauge theories one can proceed by duality transformations
from a description of the observables in terms of
gauge fields of ``electric type'' to a ``magnetic'' description
and vice--versa.
By these transformations one changes also the gauge
group and the number of gauge particles,
converts strongly coupled theories into weakly coupled ones,
confinement into screening etc.
So  it seems that none of these concepts has an intrinsic
meaning. As N. Seiberg has put it \cite{Se}: Gauge symmetries might not
be fundamental. They might only appear as long distance
artifacts of our description of the theory.

We do not subscribe to this conclusion, but
there is a lesson to be drawn from these examples which,
as a matter of fact, is not entirely new:
it makes no sense to attribute a physical interpretation
to the unobservable gauge fields of a theory.
Even the statement that some observable can be represented by a combination
of certain gauge fields may be physically meaningless.
There may be another representation of this observable in
terms of other fields in another formulation
of the theory. Thus, if it comes to the physical interpretation
of a theory, the observables have to speak for themselves.
(We mention as an aside that this point of view has been advocated
in the algebraic approach to quantum field theory for
more than three decades \cite{Ha}.) In particular,
the question of whether quarks, gluons and colour are to be
regarded as elements of physical reality has to be decided
on the basis of the observables, that is,
without recourse to gauge fields.

If one adopts this point of view, many theories seem
to lose the basis for their interpretation,
a prominent example being massless quantum electrodynamics
in two dimensions (Schwinger model).
It is well known that the local observables
in this model can be represented as functions of the free
neutral massive scalar field \cite{LoSw}.
The absence of  charged particles
from the physical sector
has been interpreted as confinement \cite{BeJo} or
screening \cite{It}, to mention only
two articles. But if one forgets about the way how the theory has
been constructed and looks only at the final result,
the observables, there seems to be no basis for either
one of these interpretations. For one could have started just
as well with a free field Lagrangian and would have
arrived at the same set of observables.
Similar examples of local gauge theories which
can be reinterpreted in terms of non--gauge theories
exist also in three dimensions \cite{Lu}.

It is the aim of the present article to show
that in spite of these ambiguities one can decide
in a precise and intrinsic manner whether or not a theory describes
at small scales a structure corresponding to the physical
idea of confined quarks and gluons,
carrying a colour charge. The particles and charges appearing
in our approach at the lower (ultraviolet) end of the
spatio--temporal scale
do not require a description in terms of unphysical degrees
of freedom  in some gauge theory. As we shall see, they can be described
as physical states and symmetries of the algebra
generated by the observables at arbitrarily small scales.
In order to avoid from the outset any confusion with
standard terminology we call these entities
{\em ultraparticles\/} \cite{Bu1} and {\em ultrasymmetries\/}.

The basic idea of our approach is to apply a new type
of renormalization group (scaling) transformation
\cite{BuVe1} to the local observables of the theory
and to proceed to the scaling limit. This method is
outlined in Sec.2.
The resulting scaling limit theory has
in general a different particle content
and different
symmetries.
These properties can be deduced from the corresponding
observables by algebraic methods developed in \cite{DoRo}.
The main results of that work
are also briefly reviewed in Sec.2.
In accordance  with the construction, the particle content
and the symmetry group of the scaling limit theory
are regarded as the ultraparticles and ultrasymmetries of
the underlying theory. Some relevant possibilities for these structures
are discussed in Sec.3, where also a physically meaningful
criterion for confinement of ultraparticles is stated.
In Sec.4 the general method is illustrated on the
example of the Schwinger
model which turns out
to have a non--trivial ultraparticle
and ultrasymmetry structure.
The article closes with a brief summary of the general method.
\section
{Ultraparticles and ultrasymmetries }
\setcounter{equation}{0}

Given a theory by specification of a Lagrangian, say, one is
faced with two basic problems. First one has to perform
the construction of the theory
and then to provide the physical interpretation
of the mathematical results. We are dealing here with the
latter problem and therefore may assume that on
the constructive side
everything has been said and done:
all correlation functions (vacuum expectation values)
of the local observables have been computed and
-- if necessary -- continued from Euclidean to
Minkowski space points. Applying the reconstruction
theorem \cite{Ha,StWi}, one can then represent the observable
fields, currents, stress energy tensor etc. as
operators on a physical Hilbert space. Our problem
is thus to extract from these data the physical
interpretation of the theory at very small scales.

Let $\phi(x)$ be a (hermitean) operator representing an
observable at the spacetime point $x$.\footnote
{\ Strictly speaking, $\phi(x)$ is an operator valued distribution,
i.e., becomes an operator only after integration with a suitable
test function.  Tensor indices etc.\ will be suppressed.}
Since we are interested in the interpretation of the theory at
small scales we have to study the effect of
a change of the spatio--temporal scale on the observables,
say $\phi$, while keeping
the scales $c$ of velocity and $\hbar$ of action fixed.
According  to the basic ideas of renormalization group theory,
(cf. \cite{Zi} and references quoted there), the observables corresponding
to $\phi$ at other scales are obtained by setting
\begin{equation}
\phi_\lambda(x)\doteq N_\lambda \, \phi(\lambda x),\ x\in {\RR}^4,
\label{2.1}
\end{equation}
where $\lambda > 0$ is a scaling factor.
We call $\phi_\lambda$ the observable at scale $\lambda$.
Whereas the action of the scaling transformation on the
argument of $\phi$ needs no explanation, its effect on
the scale of $\phi$, given by the
positive factor $N_\lambda$, is more subtle.
The familiar idea is to adjust this factor in such a way that
the expectation values of the observables at scale $\lambda$
have the same order of magnitude for all $\lambda > 0$.
The precise way in which this idea is implemented is a matter
of convention.  One may pick for example a suitable
(real) test function $f$ and consider
\begin{equation}
\phi_\lambda(f)\doteq \int d^{\,4} x\ f(x)\phi_\lambda(x).
\end{equation}
The factor $N_\lambda$ can then be fixed by demanding that
\begin{equation}
\langle V| \phi_\lambda(f)\phi_\lambda(f)|V\rangle
= \hbox{const\ \ for\ \ $\lambda>0,$}
\label{2.3}
\end{equation}
where $|V\rangle$ denotes the vacuum vector. But one
could impose this constraint just as well on some
higher moment of $\phi_\lambda(f),$ giving a
somewhat different result for $N_\lambda$ in general.
We denote the factor $N_\lambda$ which has been
fixed by imposing such a renormalization condition
by $Z_\lambda.$

If one wants to proceed to the scaling limit $\lambda \to 0$
there appears, however, a well--known
problem. The product of operators at neighbouring
spacetime points is quite singular, so the factors $Z_\lambda$
tend to 0 as $\lambda\to 0.$ One needs rather precise
information on the short distance properties of the
correlation functions and the way how $Z_\lambda$
approaches 0 in order to be able to control the
scaling limit.

This problem can be solved under favourable circumstances
(asymptotically free theories) since renormalization
group equations and perturbative methods provide
reliable information. The approach works also in certain
renormalizable theories where the underlying renormalization
group equations have a non--vanishing but small ultraviolet
fixed point. Yet in the case of theories without a (small)
ultraviolet fixed point the method does not lead to
reliable results, nor can a rigorous treatment of
non--renormalizable theories be even addressed. Moreover,
within the present context this method is also not appropriate
because of conceptual reasons. For renormalization group
equations are frequently formulated in terms of non--observable
fields which we want to omit from our discussion. Thus,
at first sight, a general definition of the scaling limit
of the observables of a theory would seem impossible.

There is, however, a solution to this problem which is
surprisingly simple \cite{BuVe1}. Within the present setting\footnote
{\ The framework in \cite{BuVe1} is more general. It applies to
arbitrary finitely localized observables, including
Wilson loops, finite strings etc.}
it can be described as follows. In a first step one
proceeds from
the unbounded operators $\phi_\lambda(f)$ to corresponding
bounded operators, such as the unitary (generating)
operators $\exp(i\phi_\lambda(f)).$ This has the effect
that, irrespective of the choice of the factor $N_\lambda$
in the defining equation
(\ref{2.1})
of $\phi_\lambda$,
there do not appear any divergence problems. The
resulting operators are bounded in norm, uniformly in
$\lambda>0$.

The second step is to restrict the four--momentum
of these bounded operators in accord with the uncertainty
principle. Roughly speaking,
one considers only observables at scale $\lambda$
which can transfer to physical states four--momentum proportional
to $\lambda^{-1}$.
Thus these observables occupy, for all
$\lambda >0$, a fixed phase space volume.
The desired operators are obtained by spacetime averages,
\begin{equation}
A_\lambda\doteq
\int d^{\,4} y\ g(y)\exp\big(i\phi_\lambda(f_y)\big),
\quad
\lambda>0,
\label{2.4}
\end{equation}
where $g$
is any test function and $f_y(x) \doteq f(x-y)$.

The results in \cite{BuVe1}
imply that this restriction on the four--momentum
transfer has in general the following effect:
if one chooses $N_\lambda$ in
(\ref{2.1})
such that the quotient
$N_\lambda/Z_\lambda$
tends to infinity in the limit of small $\lambda$,
then all correlation functions involving the corresponding
sequence of operators $A_\lambda$
converge to 0. Similarly, if
$N_\lambda/Z_\lambda$
tends to 0, then
$A_\lambda$
converges (in the sense of correlation functions)
to
$\int d^{\, 4} x\,g(x) \cdot 1$.
Thus in either case the operators
$A_\lambda$
tend in the scaling limit to multiples
of the identity. Only in the special case where the asymptotic
behaviour of
$N_\lambda$
coincides with that of
$Z_\lambda$
does it normally happen that the correlation
functions of the operators
$A_\lambda$
retain a non--trivial operator content in the scaling limit.

In view of this fact one need not know
the behaviour of
$Z_\lambda$
and may admit in the above construction all possible
factors
$N_\lambda$.
The theory takes care by itself of those choices which are unreasonable and
lets them disappear (become trivial) in the scaling
limit. It is only if
$N_\lambda$
has the ``right'' asymptotic behaviour that the sequences
$A_\lambda$
give rise to non--trivial operator limits.
One may view this method as an implicit way of introducing renormalization
group transformations.
It allows one to study
the short distance properties of local observables
in an intrinsic, model independent manner.
Actually, this method is also useful in concrete computations,
as we shall see in Sec.4.

It is mathematically convenient to regard
the operators
$A_\lambda$,
obtained by the above procedure,
as operator valued functions of the scaling parameter
$\lambda>0$.
These functions form in an obvious way an algebra\footnote
{\ Sums, products and hermitean conjugates
are pointwise defined,
$(A+B)_\lambda\doteq
A_\lambda+B_\lambda, \\
(A\cdot B)_\lambda\doteq
A_\lambda\cdot B_\lambda,~
(A^*)_\lambda\doteq
(A)_\lambda^*.
$
There exists also a norm on this algebra \cite{BuVe1}.}, called scaling algebra
in \cite{BuVe1}. For the construction of the scaling limit
of the observables one considers all expectation
values of sums and products of these operator functions
in the limit $\lambda\to 0$,
\begin{equation}
\lim_{\lambda\to 0}\
\langle V|\sum
A_\lambda A'_\lambda \cdots A''_\lambda|V\rangle
=
\langle V_0|\sum
A_0 A'_0 \cdots A''_0|V_0\rangle.
\label{2.5}
\end{equation}
Actually, these limits may only exist for suitable subsequences
of the scaling parameter $\lambda$.
We disregard this problem here and refer to the thorough
discussion of this point in \cite{BuVe1},
cf.\ also the remarks in Sec.4.
What is of interest here is the fact \cite{BuVe1}
that the limits of these expectation values determine,
by an application of the reconstruction theorem
to the scaling algebra, a local, Poincar\'e covariant theory
with vacuum vector $|V_0\rangle$.
This explains the notation on the right hand side of relation
(\ref{2.5}).
Disregarding certain exceptional cases (cf.\ the classification
of scaling limits in
\cite{BuVe1}) one can represent the operators
$A_0$
in this expression in the form
\begin{equation}
A_0=\int d^{\, 4} y\ g(y)\exp\big(i~
\hbox{const\ }\phi_0(f_y)\big),
\label{2.6}
\end{equation}
where $\phi_0$
is the scaling limit of
$\phi_\lambda$
and the value of the constant depends on the choice of the normalization
factors
$N_\lambda$.
Thus, by this universal method, one can define the scaling limit
of the observables in any given theory.

We emphasize that the operators
$A_0$
will in general have properties which are very different from those of the
original observables. They are the observables in a
``new theory'', the scaling limit theory, acting on their
own ``new Hilbert space''.
To a certain extent the situation may be compared with collision
theory, where one proceeds by the LSZ--limit from interacting
fields to free asymptotic fields.
But whereas the asymptotic fields describe the same global
(unbroken) symmetries and particle spectrum as the underlying
interacting fields, these features may change
if one proceeds to the scaling limit. As we shall see, it is this
very fact
which allows one to describe the phenomenon of confinement.

The determination of the particle spectrum
and the global symmetries of the scaling limit theory requires,
however, some further steps. So far we know (in principle)
only the vacuum expectation values of all observables
in the scaling limit theory.
Thus the Hilbert space obtained from these data by the reconstruction
theorem contains only neutral states.
There may be also charged states in the scaling limit
theory. How does one find them?

This question can be compared to the following familiar problem in group
theory. Consider a group, such as $SL(2,\CC)$,
in its defining representation and regard this representation
as a faithful picture of some ``abstract group'',
defined by its multiplication table.
How does one find the other representations of this abstract group?
There is no easy general answer to this question,
but there are general strategies which eventually lead to a solution.

The situation is very similar in our field theoretic problem.
Instead of dealing with a group we are dealing with an algebra
$\CA_{\, 0}$
which is generated by all sums and products
of the local observables
$A_0$
in the scaling limit theory and acts on the Hilbert space
of neutral scaling limit states.
We regard
$\CA_{\, 0}$
as a faithful representation of some ``abstract algebra'',
denoted by the same symbol, which is defined through
the algebraic relations in
$\CA_{\, 0}$ (such as
commutation relations, fusion rules, equations of motion etc.).
If there exist charged physical states in the scaling limit
theory they correspond to vectors in other
(disjoint) Hilbert space representations of
$\CA_{\, 0}$.
Thus the problem at hand is to determine these representations.

In analogy to the case of non--compact groups,
the algebra
$\CA_{\, 0}$
has in general a tremendous number of representations,
corresponding to very different physical situations
(states). Some of them may be very interesting, such as thermal
equilibrium states,
but they are of no relevance in the present context.
Others may be regarded from the outset as pathological,
such as states with a rapidly increasing energy density
at infinity. Thus, what is required are certain
{\em a priori conditions\/} which characterize states describing
(charged) elementary systems. Such selection criteria have been
formulated in the literature in mathematically precise terms,
cf.\ for example
\cite{Ha}. It suffices therefore to state them
here in a somewhat informal way.

The first condition which we impose on the representations
of $\CA_{\, 0}$
to be considered here is the requirement that energy--momentum
operators can be defined on the respective Hilbert spaces
and satisfy the relativistic spectrum condition
(positivity of energy in all Lorentz frames).
Note that
$\CA_{\, 0}$
contains only observables which are localized in bounded spacetime regions.
Since the energy--momentum operators
are obtained by integrating the stress energy tensor over all
of space this condition imposes
a non--trivial constraint.

The second condition implements the idea that
elementary systems are localized excitations of the vacuum.
In order to test whether a vector
$|E_0\rangle$
describes such an excitation one has to compare
the expectation values
\mbox{$\langle E_0 | A_0 | E_0\rangle$}
and the vacuum expectation values
$\langle V_0 | A_0 | V_0\rangle$.
These expectation values should coincide for all observables
$A_0$
which are localized in the spacelike complement
of any sufficiently large region, containing the localization
region of
$| E_0\rangle$.
Note that the localization properties of
$A_0$
can be prescribed by choosing in relation
(\ref{2.4})
test functions with specific support properties.

The two conditions in this selection criterion
have a clear--cut
physical interpretation,
but they provide only a rather
implicit description of the relevant set
of states and representations.
It is therefore natural to ask whether there is a more explicit
characterization. An answer to this question is given
by the following deep result of Doplicher and Roberts
\cite{DoRo} which we formulate here for the algebra
of observables in the scaling limit:
(a) the algebra
$\CA_{\, 0}$
can be extended to some field algebra
$\CF^{\,0}$,
$\CA_{\, 0}\subset \CF^{\,0}$,
which is generated by charge carrying Bose respectively Fermi
fields. (b) There is a compact group $G_0$
which acts on
$\CF^{\,0}$
by automorphisms, that is the
basic fields in
$\CF^{\,0}$
transform like vectors
in irreducible unitary representations of
$G_0$.
Moreover, the elements of
$\CA_{\, 0}$
are exactly the fixed points in
$\CF^{\,0}$
under the action of
$G_0$.
(c) The Hilbert space spanned by all state vectors
$| E_0\rangle$,
described in the selection criterion, is obtained by applying
all elements of
$\CF^{\,0}$
to the vacuum vector
$| V_0\rangle$.
(d) The algebra
$\CF^{\,0}$
and the group
$G_0$
are uniquely fixed by these properties.

This result is of importance for the problem at hand. For it
shows that
one can determine from the scaling limit
algebra
$\CA_{\, 0}$
on the Hilbert space of neutral states the global gauge group
$G_0$
of the scaling limit theory.
These are the symmetries appearing in the original theory
at very small scales, so they are the
{\em ultrasymmetries\/}.
Similarly, by analyzing the
discrete (atomic) spectrum of the mass operator on the
Hilbert space spanned by the set of vectors
$\CF^{\,0} | V_0 \rangle$
one can determine the set
$\Sigma_0$ of charged and neutral particles
in the scaling limit theory.
These are the {\em ultraparticles.\/}
In view of the canonical construction
of the scaling limit and point (d) mentioned above,
both, the ultraparticles and the ultrasymmetries
are intrinsic features of the underlying theory.

By the same general method one can recover from the algebra
$\CA$,
which is generated by the observables in the underlying theory, the global
symmetry group $G$, the algebra of charge carrying fields
$\CF$
and the physical particle spectrum $\Sigma$.

We emphasize that these structures are fixed by the class
of states which one has selected. Yet the conditions given
above may sometimes be too restrictive.
If, for example, there appear particles in the theory carrying
a gauge or quantum topological charge in the sense of
\cite{BuFr1},
one has to consider also string--like excitations of the vacuum which
are localizable in arbitrary spacelike cones.
In theories where all physical particles are massive,
these are the worst possible localization properties
of elementary states \cite{BuFr1}.
We note that the results of Doplicher and Roberts can be generalized to
this extended class
of states \cite{DoRo}. If one wants to consider
also particles carrying
electric charge one has to proceed to an even larger class
of states \cite{Bu2} and can no longer rely on the characterization
of particle states by discrete contributions in the mass spectrum
(infraparticle problem) \cite{Bu3}.

These more general cases arise whenever a particle
carrying a gauge charge appears in the physical spectrum,
i.e., is not confined. But then there are no problems
with its physical interpretation.
We will therefore restrict our attention
here to the simple case where all states of interest
on $\CA$ and
$\CA_{\, 0}$
are well--localized, the corresponding field algebras
$\CF$ and $\CF^{\,0}$
are hence generated by local field operators,
and particles correspond to sharp eigenvalues of the mass
operator. The extension of our method to the more general situation
indicated above is outlined
in Sec.4 on the example of the Schwinger model.

Before we turn to the discussion of the physical interpretation
of these structures let us comment on the problem
of computing in a given theory
the states and representations of interest, described above.
As in the case of group theory,
there are no simple general solutions,
but there are strategies.

The first approach is based on the observation
that the desired state vectors
$| E_0\rangle$
give rise to states (expectation functionals)
$\CE_0 ( \cdot ) \doteq \langle E_0 | \cdot | E_0\rangle$
on the algebra
$\CA_{\, 0}$
with certain characteristic properties.
Conversely, given a state $\CE_0 ( \cdot )$ with such  properties,
one can reconstruct the corresponding vector
$| E_0\rangle$
and representation of
$\CA_{\, 0}$
by the reconstruction theorem.
Making use of the results of
Doplicher and Roberts mentioned above one knows from
the outset that the states
$\langle E_0 | \cdot | E_0\rangle$
can be approximated by suitable sequences
of neutral states. Roughly speaking,
one has to exhibit vectors describing bilocalized excitations
of the vacuum state, where in one region about the
spacetime point 0, say, sits a charge and
in another region about the point $x$ some compensating charge,
$| E_0\times\bar E_0(x)\rangle$
in symbolic notation.
These vectors lie in the given Hilbert space of neutral states.
One then shifts $x$
to spacelike infinity with the result that the compensating
charge has no longer any effect on the local observables,
\begin{equation}
\lim_x\
\langle
E_0\times\bar E_0(x)
 | A_0 |
E_0\times\bar E_0(x)\rangle
=
\langle
E_0 | A_0 | E_0\rangle,~
A_0\in\CA_{\, 0}.
\label{2.7}
\end{equation}
Thus, by shifting a compensating charge
``behind the moon''
\cite{HaKa}, one can proceed from neutral states to charged
states and leave the Hilbert space.
The trick is to require convergence only in the weak sense
of relation
(\ref{2.7}).

The second approach works only in very special cases which may,
however, be of particular interest in physics.
Namely, let the given theory describe states and observables
which, at small scales, behave as if there
is no interaction (in a loose sense, there
is ``asymptotic freedom'').
This feature should imply that the algebra
$\CF^{\,0}$
is generated by free fields
and
$\CA_{\, 0}$
contains the $G_0$--invariant
combinations of these fields.
Moreover, basic observables in $\CA_{\, 0}$
such as the scaling limits
of observable currents, the stress energy tensor
etc.\ should be bilinear expressions
in these free fields. But from such operators one can
reconstruct the underlying fields
directly by a simple algebraic method
invented in \cite{SchL}.
\section{Structure of the scaling limit}
\setcounter{equation}{0}
We have established in the preceding section a general
method for the analysis and interpretation of a given theory at
arbitrarily small scales. Starting from the algebra $\CA$ of
local observables one first proceeds to the
corresponding algebra $\CA_{\, 0}$ in the
scaling limit theory and then determines the associated
field algebra $\CF^{\, 0}$,
the group $G_0$ of ultrasymmetries
and the set $\Sigma_0$ of ultraparticles.
The structure of $\CF^{\, 0}, G_0, \Sigma_0$
depends on detailed properties of the underlying theory
and has to be computed in each case.

In analogy to the case of collision theory,
where the interpretation of the theory is accomplished by
comparing the algebras of incoming and outgoing
fields which are related by the scattering matrix,
some interesting dynamical information is
obtained by comparing the algebras $\CA$
and $\CA_{\, 0}$, respectively the corresponding data $\CF, G, \Sigma$
and $\CF^{\, 0}, G_0, \Sigma_0$.
We discuss in the following some relevant cases.

A first important possibility
is the case $G_0 \supset G$.
It appears if there is an enhancement of symmetries
at small scales,
as it is expected in some asymptotically free gauge theories.
But we emphasize that
one cannot decide on the basis of the observables $\CA_0$
whether the extra symmetries appearing in $G_0$ are
related to some underlying local gauge group.
For these observables test the properties of the underlying
theory in an infinitesimal neighbourhood of a given spacetime point.
But, as is intuitively clear, in
such a ``neighbourhood'' one  cannot distinguish between
local and global gauge transformations.

Another interesting possibility is $G_0 \subset G$
or that there do not appear any ultrasymmetries at all
in the scaling limit.
The latter case may be met for example in theories
without ultraviolet fixed point, should they exist in
some sense as continuum theories \cite{Kl}.
One would  expect in such theories
that no choice of the
normalization factor $N_\lambda$ in relation (2.1)
leads to non--trivial operators in the scaling limit.
Then the algebra $\CA_{\, 0}$ consists of multiples
of the identity and $G_0$ is trivial.
(According to the terminology of \cite{BuVe1} the scaling limit
would be ``classical''.)

There may be also cases where the relation between $G$
and $G_0$ is more complicated. Parts of the symmetries
in $G$ may disappear in the scaling limit by a mechanism
as just described and new symmetries may appear.
Then $G$ and $G_0$ have at best some subgroup in common.

Let us now turn to the particle aspects in the scaling
limit. Here we confine ourselves to the most
interesting case of dilation--invariant scaling limit
theories. They arise if the limits in (2.5) are unique,
cf.\ the corresponding results
in \cite{BuVe1}.
In case of dilation invariance all ultraparticles in $\Sigma_0$
are massless. Moreover, the ultraparticles
do not interact at very small scales, that is the respective
incoming and outgoing collision states,
determined by standard collision theory
from the fields in $\CF^{\, 0}$, coincide \cite{BuFr2}.

Of particular interest is
the special case where the scaling limit
theory has a complete ultraparticle interpretation (``asymptotic
completeness'' in the ultra--violet). Then $\CF^{\, 0}$
describes a set $\Sigma_0$ of non--interacting massless
particles and consequently should be generated by massless free
fields.\footnote{\ In \cite{20} this fact has been established for
the case where $\CF^{\, 0}$ is generated by some local scalar field,
cf.\ also \cite{Ba} and \cite{Bu}.  But the methods can be extended to
more general situations.}
We say in this case the underlying theory is
{\em a free theory at small scales}.
(It seems likely that this definition agrees with
the notion of asymptotic freedom in renormalization group theory
\cite{Zi}.) Thus
we arrive at the following conclusion:

\vspace{3mm}
\noindent
{\it A theory has
a complete ultraparticle
interpretation if and only if it is
a free theory at small scales.}
\vspace{3mm}

The extreme opposite case is present if $\Sigma_0$ is
empty but $\CF^{\, 0}$ is non--trivial. Then the observations
made at small scales have to be interpreted without reference
to a particle picture.  This situation
is expected to occur in theories with a non--trivial
ultraviolet fixed point. There may be also intermediate cases where
$\Sigma_0$ is non--empty, but there are vectors in the space
$\CF^{\, 0} | V_0 \rangle$ which do
not have a particle
interpretation. In any case, the structure of $\Sigma_0$ is --
in the same way as that of $G_0$ -- an observable feature of the
underlying theory.

Let us finally turn to the issue of confinement of
ultraparticles. We want to cast the physical idea, that there
are ultraparticles which cannot be created by physical
operations at finite scales,
into a precise definition. To this end we
consider the scaling limit $\CF_0$ of the algebra $\CF$ of charged
physical fields. It can be computed by the same methods as outlined
above for the observables. The algebra
$\CF_0$ is an extension of the algebra $\CA_{\, 0}$
contained in $\CF^{\, 0}$, $\CF_0 \subset \CF^{\, 0}$, and
this inclusion may be proper. We say:

\noindent
{\it An ultraparticle is confined if all of its
state vectors are orthogonal to the set of vectors
$\CF_0|V_0\rangle$.\/}

Indeed, such an ultraparticle cannot be created by the
underlying fields in $\CF$  at the relevant scale
and therefore does not appear in the ``physical spectrum.''
An interesting special case occurs if the elements of $\CF_0$
are pointwise invariant under some subgroup $C_0 \subset G_0$.
Then we have:

\vspace{\smallskipamount}
\noindent
{\it Any ultraparticle carrying a non--trivial charge with
regard to $C_0$ is confined.}

\vspace{\smallskipamount}
Note that our definition of confinement of ultraparticles
does not refer to any dynamical properties.
If $\CF$ is generated by local operators (what we assume
here) the condition can be tested in arbitrarily small spacetime
regions.

We have reached now a point where we can try to answer the question
raised in the title. To this end let us recall
some basic facts of
hadronic physics.
The spectrum of hadrons, the scaling behaviour of collision cross
sections,
the jet structure of final states in collision processes etc.\ seem
to indicate that the structure of hadronic matter at small scales
can be understood in terms of particle--like entities.
Without alluding to any
particular model, let us call them quarks and gluons.

On the theoretical side these experimental facts seem to
call for a description in terms of a set $\Sigma_0$
of ultraparticles. So
the theory one is looking for has to be
a free theory at small scales.
Taking also into account the charges and
statistics of hadrons, decay rates of mesons,
ratios of total cross sections etc., one is led to
the idea that the
group of ultrasymmetries $G_0$ of the theory
contains
a ``flavour group'' $F_0$ and a ``colour group'' $C_0$.
In particular, quarks and gluons  carry non--trivial charges
with regard to $C_0$. On the other hand,
states carrying a colour charge apparently
cannot be created by physical operations. So
quarks and gluons ought to
be confined in the desired theory.

Thus a reasonable
theory of hadronic physics should have the following general structure
in the scaling limit:
the algebra $\CF^{\, 0}$ should be generated by $G_0$--multiplets
of free massless Dirac and electromagnetic fields,
describing the ultraparticles quarks and gluons.
In order that these ultraparticles are confined, the
scaling limit $\CF_0$ of the algebra of charged physical fields
$\CF$ should be pointwise invariant under the action of $C_0$.

The latter condition imposes a non--trivial dynamical
constraint on the theory in demand, but there
are good reasons to believe that a solution to this
problem is already known:
one regards the fields generating $\CF^{\, 0}$ as the
kinematical field content of the underlying theory and
``gauges'' the colour group $C_0$ by introducing corresponding
(self)couplings between the electromagnetic and Dirac fields in
$\CF^{\, 0}$. The result is of course quantum chromodynamics.
It is an asymptotically free theory
and one may therefore hope that one
can recover from the observables in the scaling limit the
structure which one has taken as a basis in its heuristic
derivation. Admittedly,  this is fiction at present since the
theory has not yet been rigorously constructed. But as all
one needs to know are the short distance properties of the theory
it suffices to construct it in a finite volume, no matter how small.
Since there is progress in this respect \cite{MaRiSe} it may well be
possible to decide on this question in the foreseeable future.

If quantum chromodynamics turns out to have an ultraparticle and
ultrasymmetry structure as outlined above, then quarks,
gluons and the colour symmetry are described by the theory as physical
facts and the same holds true for the notion of confinement.
As these features can entirely be recovered from the
observables
they cannot be ``gauged away'',
not even by duality transformations.

The theoretical interpretation of these structures in terms of gauge
fields may be ambiguous, however.
As the example of the Schwinger  model,
discussed in the subsequent section, and the more recently
discovered gauge theories show, notions such as
``strong colour--electric forces'' can be theoretical fictions
whithout any intrinsic meaning.
One can change the theoretical explanations of the
observable features of these models
in a radical, almost contradictory way.
On the other hand, quarks, gluons and the colour group seem to
be elements of physical reality which admit, on the
theoretical side, an unambiguous description in terms of
ultraparticles and ultrasymmetries.
\section{The Schwinger model at small scales}
\setcounter{equation}{0}

We have presented our approach to the short distance analysis of
quantum field theories in a  general setting in order to
make clear that the method has the character of a universal algorithm.
In this section we apply this method to the simplest model which
seems to cause some problems as far as its physical interpretation is
concerned, the Schwinger model. What is the ultraparticle and
ultrasymmetry content of this theory? The answer may seem obvious, but
there are a few surprises.

As has been mentioned in the Introduction, the algebra $\CA$ of local
observables in
the Schwinger model is generated by the free scalar field
$\phi$ in two spacetime dimensions with mass $m>0$
\cite{LoSw}. Actually, the algebra $\CA$ has also a center. But we      may
disregard this center here since it remains unaffected under scaling
transformations. More concretely, we pick any one of the pure vacuum
states of the Schwinger model and restrict the observables to the
corresponding vacuum subspace. The elements of the center then become
multiples of the identity and the vacuum can be identified with the
vacuum vector $|V\rangle$ of the free field $\phi$.

We first note that there are no charged states in the model, that is
states satisfying the selection criteria described in
Sec.2, which do not lie in the vacuum Hilbert space. This
fact can be extracted
for example from the results in \cite{FrMoSt}.
So the global symmetry group $G$ in the Schwinger model is
trivial and there holds $\CF = \CA$.

Next, let us turn to the computation of the scaling limit of the
observables. As all observables are functions of $\phi$, it suffices
to consider the operators (cf.\ relation (2.4))
\begin{equation}
A_\lambda \doteq
\int d^{\, 2} y \  g(y) \exp \left(i \int d^{\, 2} x \  f(x-y) N_\lambda
\phi(\lambda x)\right), \quad \lambda > 0,
\label{4.1}
\end{equation}
where $f$, $g$ are arbitrary real test functions and $N_\lambda$ is an
arbitrary positive normalization factor. We have to compute the corresponding
expectation values
$\langle V|A_\lambda^{(1)} A_\lambda^{(2)} \cdots A_\lambda^{(n)}|V\rangle$,
where $A_\lambda^{(j)}$ has the same form as $A_\lambda$ with
$f$, $g$, $N_\lambda$ replaced by arbitrary test functions and
normalization factors $f^{(j)}$, $g^{(j)}$, $N_\lambda^{(j)}$,
and to proceed to the limit $\lambda \to 0$.

In view of the $c$--number commutation relations of the free field
$\phi$ there holds
\begin{eqnarray}
A_\lambda^{(1)} A_\lambda^{(2)} \cdots A_\lambda^{(n)}
&=&\int \cdots \int \prod_{j=1}^n d^{\, 2} y_j \  g^{(j)} (y_j)
\, \eta_\lambda
(y_1,\ldots y_n) \times
\label{4.2}
\\
&&\times \exp \left(i\int d^{\, 2} x \
\Bigl(\sum_{k=1}^n N_\lambda^{(k)} f^{(k)} (x-y_k)\Bigr)\phi(\lambda x)
\right),
\nonumber
\end{eqnarray}
where $\eta_\lambda(y_1,\ldots y_n)$ is a phase factor given by
\begin{eqnarray} \label{4.3}  & &
{\quad} \eta_\lambda(y_1,\ldots y_n) =
\\
& & =
\exp\left(2 \pi i \, {\rm Im} \sum_{k>l} N_\lambda^{(k)} N_\lambda^{(l)}
\int {d \bp \over 2 \sqrt{\bp^2+\lambda^2 m^2} }
\overline{\widetilde{ f^{(k)}} (p {(\lambda)})} \widetilde{ f^{(l)}}
(p {(\lambda)}) e^{ip {(\lambda)} (y_l-y_k)}\right).
\nonumber
\end{eqnarray}
Here $\tilde f$ denotes the Fourier transform of $f$ and
$p {(\lambda)} \doteq (\sqrt{\bp^2+\lambda^2 m^2},\bp )$.
Thus the vacuum expectation value of the operator
(\ref{4.2}) is given by
\begin{eqnarray}
\, \,  \langle V|A_\lambda^{(1)} A_\lambda^{(2)}
\cdots A_\lambda^{(n)} |V\rangle
= \int\cdots\int \prod_{j=1}^n d^{\, 2} y_j \  g^{(j)} (y_j) \,
\eta_\lambda(y_1, \dots y_n) \times
&&
\label{4.4}
\\
\times \exp\left(-\pi \int
{d\bp \over 2 \sqrt{\bp^2+\lambda^2 m^2}} \, \Bigr| \sum_{k=1}^n
N_\lambda^{(k)} e^{ip {(\lambda)} y_k}
\widetilde{f^{(k)}} (p {(\lambda)})
\Bigr|^2\right).
&&
\nonumber
\end{eqnarray}

\noindent Let us analyze the possible behaviour of this expression for
$\lambda \to 0$.

(a) If there is some $N_\lambda^{(k)}$ which diverges as $\lambda \to 0$
 the expression vanishes, no matter how the other normalization factors
behave in this limit. This is a consequence of the dominated
convergence theorem, applied to the $y$--integrations, and the fact that
the exponential function in (\ref{4.4}) vanishes for almost any choice of
$y_1, \ldots y_n$.

(b) Next let us assume that all $N_\lambda^{(k)}$ converge as
$\lambda \to 0$. Then $\eta_\lambda(y_1,\ldots y_n)$ converges, too.
Note that the logarithmic divergence appearing in (\ref{4.3}) in the
integration if $\lambda=0$ does not cause any problems since the
divergent contribution is real. The properties of the exponential
function in (\ref{4.4})
are more subtle, however. First, it is obvious that the
logarithmic divergence of the integral now becomes effective if
$\sum\limits_{k=1}^n N_0^{(k)} \widetilde{ f^{(k)}} (0) \ne 0$.
Then the expression (\ref{4.4}) vanishes in the limit $\lambda \to 0$.
If the sum is equal to $0$, one finds by an asymptotic expansion that
the exponential function in (\ref{4.4}) can be replaced in the limit
$\lambda \to 0$ by
\begin{equation}
\exp\left( -\pi \int {d\bp \over 2 |\bp |} \, \Bigl|
\sum_{k=1}^n N_0^{(k)} e^{ip {(0)} y_k} \widetilde{ f^{(k)}}
(p {(0)}) \Bigr|^2 - \pi |{\rm ln}\lambda| \Bigl|\sum_{k=1}^n
N_\lambda^{(k)}
\widetilde{ f^{(k)}}(0) \Bigr|^2 \right),
\label{4.5}
\end{equation}
where $p {(0)}=(|\bp |,\bp)$.
Because of the appearance of the divergent factor
$|{\rm ln}\lambda|$
it is apparent that also the rate of convergence of the
normalization factors $N_\lambda^{(k)}$ matters. This fact leaves room
for some interesting phenomenon, as we shall see.

If one replaces in (\ref{4.4}) the exponential function by (\ref{4.5})
and $\eta_\lambda(y_1,\ldots y_n)$ by its limit
$\eta_0(y_1,\ldots y_n)$ one finds that the resulting expression
has a very simple interpretation. To exhibit this fact, let
\begin{equation}
\hat A_0^{(k)}  \doteq \int d^{\, 2} y \, g^{(k)} (y) \,
\exp\left(
i\int d^{\, 2} x f^{(k)} (x-y) N_0^{(k)} \phi_0 (x) \right),
\label{4.6}
\end{equation}
where $\phi_0$ is the free massless scalar field, and let
$|\hat V_0\rangle$ be the corresponding vacuum vector. We recall that
the exponential functions of the smoothed--out field $\phi_0$
(Weyl--operators) can be defined on a vacuum Hilbert space with
positive metric, cf.\ also the remarks below.
Next, let $C$ be some classical random variable with
spectrum $\RR$ and let $\langle \, \cdot \, \rangle_G$
be the Gaussian state (probability measure) given by
\begin{equation}
\langle \exp(i\zeta C) \rangle _G  \doteq
e^{-\zeta^2/2}, \quad \zeta \in \RR.
\label{4.7}
\end{equation}
We set
\begin{equation}
\check{A}_\lambda^{(k)} \doteq
\exp\left( i \,
| 2 \pi {\rm ln} \lambda|^{1/2} (N_\lambda^{(k)}-N_0^{(k)})
 \widetilde{ f^{(k)}}
(0) C\right), \quad \lambda>0.
\label{4.8}
\end{equation}
It then follows from the preceding computations that, apart from
terms which vanish for $\lambda \to 0$, there holds
\begin{equation}
\langle V| A_\lambda^{(1)} \cdots A_\lambda^{(n)} | V \rangle =
\langle \hat V_0 | \hat A_0^{(1)} \cdots \hat A_0^{(n)} |
\hat V_0 \rangle \, \langle \check A_\lambda^{(1)} \cdots \check
A_\lambda^{(n)} \rangle _G
\label{4.9}
\end{equation}
if $\sum\limits_{k=1}^n N_0^{(k)} \widetilde{ f^{(k)}} (0) = 0$.
If this
sum is different from zero the right hand side of (\ref{4.9}) has to
be replaced by $0$.
Thus we see that at small scales the theory can be interpreted in
terms of a free massless scalar field and an independent classical
random variable $C$. So there appears a new
degree of freedom at small scales.

(c) Let us finally discuss the case where the normalization factors
$N_\lambda^{(k)}$ have a behaviour at $\lambda=0$ such that the
expectation values (\ref{4.4}) do not converge. Then there still holds
\begin{equation}
| \langle V| A_\lambda^{(1)} \cdots A_\lambda^{(n)} | V \rangle |
\leq \prod_{k=1}^n \int d^{\, 2} y \, |g^{(k)} (y)|,
\label{4.10}
\end{equation}
so the expectation values are uniformly bounded for any choice of
the normalization factors. It thus follows from elementary calculus
(Heine--Borel theorem) that there always exist subsequences
$\lambda_i \to 0$
such that the expectation values converge. One is then in the
framework either of case (a) or of case (b), so one gets nothing
new. The non--convergence of the expectation values simply means that
one has picked at different scales observables which test rather
different properties of the state. But by proceeding to subsequences
one can always correct this inappropriate choice.

These results illustrate the general method of the scaling algebra
\cite{BuVe1}. It is an algorithm which does not require any a priori
information about the ``correct'' normalization factor $Z_\lambda$.
As a matter of fact, if we had
plugged into relation (\ref{4.1}) the -- according to the canonical
dimension of the field $\phi$ -- obvious choice
$N_\lambda = 1$
we would have missed the interesting point that the field contains
 also a classical contribution $C$ with anomalous
dimension. By our unbiased approach this fact has come to light.

Summing up, we see that the algebra of observables $\CA$
in the Schwinger model has in the
scaling limit the form of a tensor product,
$\CA_{\, 0} = \hat \CA_{\, 0} \otimes \check \CA_{\, 0}$,
where $\hat \CA_{\, 0}$ is the algebra generated by the free massless
scalar field
$\phi_0$ in Weyl form and $\check \CA_{\, 0}$ is an abelian algebra
generated by the classical variable $C$. (Note that the center
$1 \otimes \check \CA_{\, 0}$
of $\CA_{\, 0}$ has to be distinguished from the center of the original
algebra of observables, which is still present as an additional ``dummy
variable''.) Moreover, the pure vacuum state on $\CA$
becomes a mixed state on $\CA_{\, 0}$
which may be interpreted as a vacuum state containing a
condensate of zero energy Bosons.

We mention as an aside that the latter phenomenon is a peculiarity of
the two--dimensional world. In higher dimensions the scaling limit of a
pure vacuum state is always pure \cite{BuVe1}.

Let us turn now to the determination of the ultraparticles and
ultrasymmetries in the Schwinger model. We do not aim here at a complete
description of $G_0$, $\Sigma_0$
and restrict ourselves to showing that there appear charged physical
states in the scaling limit. To some extent we can follow here ideas
already expounded in the literature, cf.\ for example \cite{StWi2}.

The center of
$\CA_{\, 0}$ is of little interest, in particular since its
appearance is a peculiarity of two dimensional theories. So we need
not think up new states on this part of the algebra and may
restrict our attention to states
$\langle E_0 | \cdot | E_0 \rangle$
on $\CA_{\, 0}$ of the form
\begin{equation}
\langle E_0 | A_0^{(1)} \cdots A_0^{(n)} | E_0 \rangle =
\langle \hat E_0 | \hat A_0^{(1)} \cdots \hat A_0^{(n)} | \hat E_0 \rangle
\, \langle \check A_0^{(1)} \cdots \check A_0^{(n)} \rangle_G,
\label{4.11}
\end{equation}
where
$\langle \hat E_0 | \cdot | \hat E_0 \rangle$
is a suitable functional on the algebra
$\hat \CA_{\, 0}$ generated by the
exponential functions of the smoothed--out
free massless field $\phi_0$.
In view of the fact that $\langle \, \cdot \, \rangle_G$ is kept fixed
 we may ignore in the following
the center
$1 \otimes \check \CA_{\, 0}$ and identify the algebra $\CA_{\, 0}$
with its free field part $\hat \CA_{\, 0}$.

The specification of functionals
$\langle \hat E_0 | \cdot | \hat E_0 \rangle$
on
$\hat \CA_{\, 0}$ is greatly simplified by the fact that the algebraic
relation (\ref{4.2}) holds also for the operators
$\hat A_0^{(k)}$ if one replaces  $N_\lambda^{(k)}$,
$\eta_\lambda (y_1,\ldots y_n)$,
$\phi(\lambda x)$ by $N_0^{(k)}$,
$\eta_0(y_1,\ldots y_n)$ and $\phi_0(x)$,
respectively. It therefore suffices to specify the functionals for a
single Weyl--operator of the form
$\hat W_0 (f) = \exp \left(i \int d^{\, 2} x \  f(x) \phi_0 (x)\right)$,
where $f$ is an arbitrary real test function.

The vacuum state on $\hat \CA_{\, 0}$ is fixed by setting
\begin{equation}
\langle \hat V_0 | \hat W_0 (f) | \hat V_0 \rangle =
\left\{ \begin{array}{c@{\quad \mbox{if} \quad}l}
\exp{ \big( -\pi \int {\mbox{{\small $d$}} \bp \over
{\mbox{{\small $2$}} |\bp |}}
| \tilde f  ( p {(0)} ) |^2 \big) }  & \tilde f(0) = 0 \\
0 & \tilde f(0) \ne 0,
\end{array} \right.
\label{4.12}
\end{equation}
where $p {(0)} = (|\bp |,\bp)$ as before.\footnote{
In view of the discontinuity at $f=0$ this vacuum state is a
so--called non-regular state. As a consequence, the
smoothed--out
field operators $\phi_0 ( f )$ cannot be defined on the corresponding
Hilbert space if $\tilde f(0) \ne 0$, in contrast to the
Weyl--operators.}
The simplest class of states on
$\hat \CA_{\, 0}$ next to the vacuum are the so--called
coherent states. They give rise to expectation values of
the form
\begin{equation}
\langle \hat E_0 | \hat W_0 (f) | \hat E_0 \rangle =
e^{iL(f)} \, \langle \hat V_0 | \hat W_0 (f) | \hat V_0 \rangle,
\label{4.13}
\end{equation}
where $L$ is a linear functional on the space of real test functions
which vanishes on all functions of the form
$f(x) = \Box g(x)$. Phrased differently,
$L$ is a weak solution of the wave equation. Picking any such $L$
 one obtains from (\ref{4.13}) with the help of
relation (\ref{4.2}) a state on $\hat \CA _0$ which, by the
reconstruction theorem, gives rise to a representation of
$\hat\CA_{\, 0}$ on
some Hilbert space. The problem is that this representation may not
describe elementary systems in the sense of the selection criterion
given in Sec.2. So one has to determine the subset of
functionals $L$ which are admissible in that sense.

Instead of giving the
complete solution of this problem it may suffice here to exhibit an
interesting class of functionals $L$ which comply with the selection
criterion. Let $\rho(\bx )$, $\bx \in \RR$, be any real, symmetric test
function with support in the interval $[-r,r]$, say, and let
$\int d\bx \, \rho ( \bx ) = q$, where $q\ne 0$
is some fixed ``charge''. We put
\begin{equation}
L(f) \doteq  \sqrt{ 2 \pi}
\int d\bp {\tilde \rho(\bp ) \over { 2 i \bp}}
\left( \tilde f (p {(0)}) - \tilde f (-p {(0)}) \right).
\label{4.14}
\end{equation}
This functional is well--defined, real and linear
and there holds $L( \Box g) = 0$.

One can show that the representation of
$\hat \CA_{\, 0}$ which is induced
by the state (\ref{4.13}) for the given
functional $L$ admits energy--momentum operators which are the
generators of space--time translations and satisfy the relativistic
spectrum condition. Since we are dealing with coherent states this
fact is almost obvious. For example, the expectation value of the
energy operator in the state fixed by $L$ is given by
$( 1 / 2 ) \int d\bx | \rho(\bx )|^2$.
A rigorous proof of the assertion is obtained by applying standard
methods, cf.\ for example \cite{Ro}.

Next we have to determine the configuration space properties
 of the functional $L$. To this end we introduce the function
\begin{equation}
\tilde h (\bp ) \doteq {1 \over 2} \left(\tilde f ( | \bp |, \bp ) +
\tilde f ( - | \bp |, \bp )\right)
\label{4.15}
\end{equation}
and note that the functional $L$ can be represented in the form
\begin{equation}
L(f) = - { \sqrt{ \pi \over 2}}
\int d\bx \Bigl(h(\bx ) - h(-\bx ) \Bigr)
\int_{-\infty}^{\bx} d\by \rho(\by )
\label{4.16}
\end{equation}
since $\rho$ is symmetric. Now let
$a = (a_0, \ba) \in \RR^2$, $a_0 > 0$,
and let $f$ be any test function which has support in the region
$\{x = (x_0, \bx ) \in \RR^2: |x_0| + |\bx - \ba | < a_0 \}$.
Then the Fourier transform $h$ of $\tilde h$ has support in the interval
$[\ba -a_0, \ba +a_0]$.
Hence if this interval does not intersect with $[-r,r]$,
we infer from (\ref{4.16}), bearing in mind that
$\int d\bx \rho (\bx ) = q$, that
\begin{equation}
L(f) = \mp \pi  q \tilde f (0)
\qbox{if} a_0 < \pm ( \ba - r).
\label{4.17}
\end{equation}
This result can be rephrased in more geometrical terms.
If one regards the interval $[-r,r]$ as a region in Minkowski space
$\RR^2$ at time $x_0 =0$, then relation (\ref{4.17}) holds if $f$ has
support in the right, respectively left spacelike complement of that region.

This fact has interesting consequences for the localization properties
of the corresponding state. Plugging the result (\ref{4.17}) into
relation (\ref{4.13}) we find that
\begin{equation}
\langle \hat E_0 | \hat W_0 (f) | \hat E_0 \rangle =
e^{\mp i \pi q \tilde f (0)} \,
\langle \hat V_0 | \hat W _0 (f) | \hat V_0 \rangle.
\label{4.18}
\end{equation}
If $\tilde f (0) \ne 0$, then
$\langle \hat V _0 | \hat W _0 (f) | \hat V_0 \rangle = 0$,
hence there holds also
$\langle \hat E_0 | \hat W_0 (f) | \hat E_0 \rangle = 0$.
On the other hand, if $\tilde f(0) = 0$ it follows from (\ref{4.18})
that $\langle \hat E_0 | \hat W_0 (f) | \hat E_0 \rangle =
\langle \hat V _0 | \hat W _ 0 (f) | \hat V _0 \rangle$.
Hence the two states
$\langle \hat E_0 | \cdot | \hat E_0 \rangle$ and
$\langle \hat V _0 | \cdot | \hat V _0 \rangle$
coincide on all operators
$\hat W_0 (f)$, provided $f$ has support in the right or left spacelike
complement of the interval $[-r,r]$ at time $x_0 = 0$.
In this sense the state
$\langle \hat E _ 0 | \cdot | \hat E _0 \rangle$
corresponds to an elementary excitation of the vacuum.

The conclusion that this excitation is
strictly localized in the region $[-r,r]$ would be
wrong, however. This can be seen if one chooses a test function $       f$ of
the form $f = f_+ + f_-$, where $f_\pm$ have support in the right and
left spacelike complement of the interval $[-r,r]$
at time $x_0 = 0$, respectively.
Because of the linearity of the functional $L$ one obtains in this case
\begin{equation}
\langle \hat E_0 | \hat W_0 (f) | \hat E_0 \rangle =
e^{-i \pi  q(\widetilde{f _+} (0) -
\widetilde{ f_-} (0))} \,
\langle \hat V_0 | \hat W_0 (f) | \hat V_0 \rangle.
\label{4.19}
\end{equation}
Picking $f$ with $\widetilde{ f _+} (0) + \widetilde{ f_-} (0) = 0$
and $\widetilde{ f _\pm} (0) \ne 0$
it is thus possible to discriminate the two functionals in the
spacelike complement of any bounded region, no matter how large.

It remains to be shown that the state
$\langle \hat E_0 | \cdot | \hat E_0 \rangle$
cannot be represented by a vector in the Hilbert space of neutral
states. This follows by standard arguments, using the fact that
the function $|\tilde \rho (\bp )|^2 / |\bp |$
is not integrable at $\bp = 0$.
Instead of giving a formal proof the following argument may be more
instructive. Consider the identically conserved ``electric'' current
$j_\mu (x) = \partial^\nu \varepsilon_{\mu\nu}  \phi_0 (x)$,
where $\partial^\nu$ denotes the space--time derivatives and
$\varepsilon_{\mu\nu}$ the antisymmetric tensor.
By a straightforward computation one finds that
\begin{equation}
\langle \hat E_0 | j_0 (x) | \hat E_0 \rangle =
{1 \over \sqrt{2 \pi}}
\int d\bp \  \tilde\rho (\bp ) \cos (|\bp | x_0)
\cos( \bp \bx )
\label{4.20}
\end{equation}
and consequently
\begin{equation}
\int_{x_0={\rm const}}d\bx  \  \langle\hat E_0 | j_0 (x) | \hat E_0 \rangle =
q.
\label{4.21}
\end{equation}
So the functional
$\langle \hat E_0 | \cdot| \hat E_0 \rangle$
describes a charged state. On the other hand, since ``Gauss' law''
$j_0 = \partial_1 \phi_0$ holds
for the current $j_\mu(x)$, all the locally generated vectors in the space
$\hat \CA_{\, 0} | \hat V_0 \rangle$
carry zero charge.

Summing up, we have shown that the algebra
$\hat \CA_{\, 0}$ admits states
describing charged elementary systems. Intuitively speaking, these
systems carry a well--localized
``electric'' charge to which an infinite string is
attached which is strongly fluctuating. One cannot decide by
measurements with certainty whether
this string emanates to plus or minus
spacelike infinity. The states look in the right and left spacelike
complement of the localization region of the charge like the vacuum,
but the presence of the string can be established if one makes
suitable coincidence measurements in both regions, cf.\
relation (\ref{4.19}). Thus these states describe the perhaps simplest
example of a quantum topological charge in the sense of \cite{BuFr1}.

Let us turn now to the determination of the charged fields $\CF^{\, 0}$,
the ultrasymmetries $G_0$ and the set $\Sigma_0$ of ultraparticles,
which are
fixed by the class of states which we have constructed. Here we
cannot rely on the general results of Doplicher and Roberts which hold
only in physical spacetime $\RR^4$. But the necessary computations are
simple enough to be carried out explicitly.

The charged fields
in $\CF^{\, 0}$ creating the vectors $| \hat E_0 \rangle$ from the
vacuum $| \hat V_0 \rangle$ are obtained by a limiting procedure from
operators in
$\hat \CA_{\, 0}$ (akin to bosonization formulas). They can formally be
represented according to
\begin{equation}
\hat \psi_\rho (x) \doteq \lim_{R\to \infty}
\exp \big( i  \phi_0 ( \rho_{R, x} ) \big),
\quad x \in \RR^2,
\label{4.22}
\end{equation}
where the Fourier transform of $\rho_{R, x}$ is given by
\begin{equation}
\widetilde{\rho_{R, x}} ( p ) = { 1 \over \sqrt{ 2 \pi}} \, p_0
\tilde \rho ( \bp ) { 1 \over \bp } \left( 1 -
{ \sin ( \bp R ) \over \bp R } \right) e^{i p x}.
\label{4.23}
\end{equation}
Note that $\rho_{R, x}$ is real and has compact support.
The rule of the game is that one first has to compute the expectation
values involving these operators for fixed $R$ and then to proceed to
the limit $R \to \infty$. This procedure corresponds to the process of
``shifting a compensating charge behind the moon'', mentioned in
{Sec.2.}

By a calculation analogous to the case of relation
(\ref{4.4}) one finds that the expectation value of any product
$\hat \psi_{\rho_1} (x_1) \cdots \hat \psi_{\rho_n} (x_n)$
of field operators in the vacuum state
$\langle \hat V_0 | \cdot | \hat V_0 \rangle$
vanishes if $q_1 + \cdots + q_n \ne 0$.
Hence the mappings
\begin{equation}
\hat \psi_\rho (x) \to
e^{i \xi q} \, \hat \psi _\rho (x), \quad
\xi \in \RR
\label{4.24}
\end{equation}
define global gauge transformations
of the fields in $\CF^{\, 0}$.
Moreover, the observables in $\CA_{\, 0}$ are invariant under these
transformations. Thus, according to
our terminology, the gauge group generated by the transformations
(\ref{4.24}) is an ultrasymmetry of the Schwinger model.

It remains to determine the ultraparticle content of the model. To
this end let us first have a look at the energy--momentum spectrum of
the vectors $| \hat E_0 \rangle = \hat \psi _\rho | \hat V_0 \rangle$
which can be extracted from the two--point function
\begin{equation}
\langle \hat V_0 | \hat \psi_\rho^* \,
\hat \psi_\rho^{} (x) | \hat V_0 \rangle
= \exp \left( -\int {d\bp \over 2|\bp |} |\tilde \rho (\bp )|^2
(1-e^{ip {(0)} x}) \right).
\label{4.25}
\end{equation}
One can show that the Fourier transform of this function does not have
a discrete ($\delta$--function) mass--shell contribution, the reason
being again the non--integrability of the function
$|\tilde\rho (\bp )|^2 / |\bp |$ at $\bp = 0$. Thus,
according to the conventional definition of particle states, the
charged states do not seem to have a particle interpretation. But this
conclusion
would be premature. In order to see this let us look more closely
at the $x$--dependence of the expectation value (\ref{4.20}) which
can easily be computed. It is given by
\begin{equation}
\langle \hat E_0 | j_0 (x) | \hat E_0 \rangle =
{1 \over 2} \Bigl(\rho (\bx + x_0 ) + \rho(\bx - x_0 ) \Bigr).
\label{4.26}
\end{equation}
Hence the charge density of the state $|\hat E_0 \rangle$
propagates with the velocity of light through spacetime and is localized
in regions of fixed size. It is therefore meaningful to say that the
state described by $| \hat E_0 \rangle$ contains a charged
massless ultraparticle. The absence of a discrete contribution in the
mass spectrum of (\ref{4.25}) is due to the fact that this particle
cannot be separated from clouds of neutral particles
accompanying it. In view of the fact that ``Gauss' law''
holds for the
charge this is not a surprise \cite{Bu3}.

Thus the charged ultraparticles
in the Schwinger model turn out to be
infraparticles
in the sense of \cite{Sch2} as well. This observation shows
that the conventional (Wigner) particle concept may not always be
sufficient to describe the ultraparticles in a theory. Yet since there
exists a systematic way to uncover all particle like structures from
the observables, including infraparticles, cf.\ \cite{BuPoSt}
and \cite{Bu1},
this point does not provide an obstacle to our general approach.

The second ultraparticle appearing in the Schwinger model is neutral and
massless. Its states can be generated from the vacuum by the observable
$\partial_\mu \phi_0(x)$.
In contrast to the charged ultraparticle  whose state vectors
are orthogonal to all vectors in
$\CF_0 | \hat V_0 \rangle = \CA_{\, 0} | \hat V_0 \rangle$,
this particle is not confined. It is what remains of the massive
particle in the scaling limit.

This completes our analysis of the observables of the Schwinger model
in the scaling limit, cf.\ \cite{BuVe2} for a more detailed account.
The results nicely illustrate the general picture of ultraparticles,
outlined in Sec.3. They show that ``electrically'' charged
ultraparticles
appear in the model at small scales which have no counterpart
at large scales, so they are confined. Within the theoretical setting
this must be regarded as an observable fact.

What seems to be somewhat arbitrary, however, is the theoretical
explanation of
this phenomenon, offered by the Schwinger model. It relies on the
picture that the electric force is strong at large
distances and binds opposite charges into massive
bound states. Since the theory is asymptotically free one may then
argue that the force is turned off at small scales
so that charged ultraparticles
can appear as physical states in the scaling limit.

This is an illustrative and consistent explanation of the observable
features of the Schwinger model, but it does not have any more
predictive power
than the explanation suggested by massive free field theory. As there
are no forces present in the latter model, that explanation looks
rather different, however.

{}From the point of view of free field theory the appearance of charged
elementary systems in the scaling limit is a pure quantum effect. In
order to exhibit this mechanism let us proceed from
the field $\phi_\lambda$ at scale $\lambda$ to
the corresponding time $x_0=0$ field $\varphi_\lambda$ and its
canonically conjugate field (time derivative) $\pi_\lambda$.
If one integrates $\pi_\lambda$ with a real test function
$g$ one finds that for any normalized state vector $|S\rangle$ in the
underlying Fock space there holds \cite{ReSi}
\begin{equation}
\langle S| \pi_\lambda(g)^2 | S \rangle \le
\int d\bp \sqrt{\bp^2+\lambda^2 m^2} \, |\tilde g(\bp )|^2 \cdot
\langle S |(2N+1)|S \rangle,
\label{4.27}
\end{equation}
where $N$ is the particle number operator. In view of the canonical
commutation relations
$[\varphi_\lambda(f), \pi_\lambda(g)] =
i \int d\bp {\tilde f (-\bp )} \tilde g (\bp ) \cdot 1$
one obtains from this estimate
according to the uncertainty principle the
inequality
\begin{eqnarray}
\langle S| \varphi_\lambda (f)^2 | S \rangle
\!\!\!\!&-&\!\!\!\!
\langle S | \varphi_\lambda (f) | S \rangle ^2
\ge
\label{4.28}
\\
&\ge&
\sup_g {| \int d\bp \, {\tilde f(-\bp )} \tilde g(\bp ) |^2 \over
4 \int d\bp \sqrt{\bp^2 + \lambda^2 m^2} \, | \tilde g (\bp)|^2 \cdot
\langle S| (2N+1)| S \rangle}
\nonumber
\\
&=&
{1 \over 4} \int {d\bp \over \sqrt{\bp^2+\lambda^2 m^2}} \,
| \tilde f (\bp )|^2 \cdot {1 \over \langle S|(2N+1)|S \rangle}.
\nonumber
\end{eqnarray}
Hence if $\tilde f(0) \ne 0$,
the fluctuations of $\varphi_\lambda (f)$ become infinitely large in
the scaling limit, which makes it impossible to discriminate different
states by these operators (the corresponding measurements would have
infinite error bars).

As a result, certain field configurations which can be distinguished
at finite scales become indistinguishable in the scaling limit. If one
adds for example to some state
$\langle S| \cdot | S \rangle$
an external field $F$ at time $x_0 = 0$
this amounts to
replacing in the original expectation values the operator
$\varphi_\lambda(f)$ by
$\varphi_\lambda(f) + F(f) \cdot 1$.
It therefore does not affect the fluctuations of
$\varphi_\lambda(f)$.
Consequently, the value of $F(f)$ becomes meaningless at small scales if
$\tilde f(0) \ne 0$. On the other hand,
for test functions $f$ with $\tilde f (0) = 0$ only the spatial
derivative $\partial_1 F$ of the field $F$
is tested. So fields $F$
differing by a constant cannot be distinguished
at small scales and kink--like field
configurations, which contain an infinite amount of energy in their
asymptotically constant pieces, look in the scaling limit like
localized excitations with finite energy. Thus quantum effects provide
an alternative explanation of the appearance of ultraparticles and
ultrasymmetries in the Schwinger model and the gauge fields may then
be viewed as nothing more but some choice of integration variables in
a functional integral.
\section{Summary}
\setcounter{equation}{0}
We have presented in this article a general method which allows one to
extract from the local observables in a given theory the structure and
interpretation of the theory at very small spatio--temporal scales.
The necessary steps are indicated in the following diagram:
$$
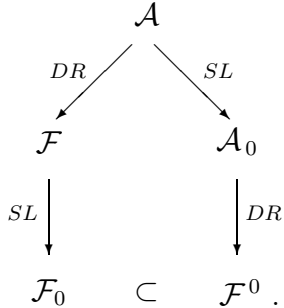
\setlength{\unitlength}{1mm}
\begin{picture}(50,50)
\put(5,0){\makebox(10,10){$\CF_0$}}
\put(18,0){\makebox(10,10){$\subset$}}
\put(31,0){\makebox(10,10){\ $\CF^{\, 0}$\ .}}
\put(4,13){\makebox(5,5){$\scriptstyle \rm SL$}}
\put(10,20){\vector(0,-1){10}}
\put(35,20){\vector(0,-1){10}}
\put(36,13){\makebox(5,5){$\scriptstyle \rm DR$}}
\put(5,20){\makebox(10,10){$\CF$}}
\put(30,20){\makebox(10,10){$\CA_{\, 0}$}}
\put(10,32){\makebox(5,5){$\scriptstyle \rm DR$}}
\put(21,38){\vector(-1,-1){10}}
\put(24,38){\vector(1,-1){10}}
\put(30,32){\makebox(5,5){$\scriptstyle \rm SL$}}
\put(18,37){\makebox(10,10){$\CA$}}
\end{picture}
$$
Here DR stands for the Doplicher--Roberts construction of charged
physical fields from local observables \cite{DoRo} and SL denotes the novel
method of computing scaling limits, invented in \cite{BuVe1}. As is
indicated by the diagram, the whole structure is completely fixed
once the algebra $\CA$ of observables is given.

The ultraparticles and ultrasymmetries appearing in the
scaling limit can be extracted from the algebra
$\CF^{\, 0}$ and are thus
an intrinsic property of the underlying theory which must be
regarded as an (in principle) observable fact. Moreover, one can
distinguish between confined and non--confined ultraparticles by
comparing the algebras
$\CF_0$ and $\CF^{\, 0}$. Loosely speaking, there is
room for the phenomenon of confinement of ultraparticles
in the theoretical setting
since the operations DR and SL do not need to commute.

These results do not rely in any way on the existence of gauge fields.
As a matter of fact, such an extension of an algebra of observables
$\CA$ to a theoretical superstructure may be
ambiguous, as one knows from various examples.

The general picture of the short distance structure of asymptotically
free, confining theories which emerges from these results has been
illustrated by
the Schwinger model. It seems feasible that with
some effort it can also be established in physically more interesting
theories, such as quantum chromodynamics.
\pagebreak

\bigbreak\medskip
\noindent
{\Large\bf Acknowledgements}

\vspace{\medskipamount}
\noindent
I enjoyed discussions with J. Gomis, H. Joos
and I. Ojima on the subject of
this paper  and I am very grateful to the members of RIMS and its director
H. Araki for their hospitality and care during my visit to Kyoto
University. I also acknowledge a research grant from the Japan Society
for the Promotion of Science (JSPS) and financial support from the
Deutsche Forschungsgemeinschaft (DFG).

\end{document}